\documentclass[preprint,aps,prd,superscriptaddress,showpacs,nofootinbib]{revtex4}%
\usepackage{graphicx}
\usepackage{amsmath}
\usepackage{amssymb}
\usepackage{bm}
\usepackage{epsfig}
\usepackage{mathrsfs}
\usepackage{hyperref}

\begin{document}
\preprint{}
\title{\mbox{}\\[10pt]
$\bm{Z}$-boson decays to a vector quarkonium plus a photon
}
\author{Geoffrey~T.~Bodwin}
\email[]{gtb@anl.gov}
\affiliation{High Energy Physics Division, Argonne National Laboratory,
Argonne, Illinois 60439, USA}
\author{Hee~Sok~Chung}
\email[]{hee.sok.chung@cern.ch}
\affiliation{Theory Department, CERN, 1211 Geneva 23, Switzerland}
\author{June-Haak~Ee}
\email[]{chodigi@gmail.com}
\affiliation{Department of Physics, Korea University, Seoul 02841, Korea}
\author{Jungil~Lee}
\email[]{jungil@korea.ac.kr}
\affiliation{Department of Physics, Korea University, Seoul 02841, Korea}
\date{\today}
\begin{abstract}

We compute the decay rates for the processes $Z\to V+\gamma$, where 
$Z$ is the $Z$ boson, $\gamma$ is the photon, and $V$
is one of the vector quarkonia $J/\psi$ or $\Upsilon(nS)$, with $n=1$,
$2$, or $3$. Our computations include corrections through relative
orders $\alpha_s$ and $v^2$ and resummations of logarithms of
$m_Z^2/m_Q^2$, to all orders in $\alpha_s$, at NLL accuracy. ($v$ is
the velocity of the heavy quark $Q$ or the heavy antiquark $\bar{Q}$
in the quarkonium rest frame, and $m_Z$ and $m_Q$ are the masses of
$Z$ and $Q$, respectively.) Our calculations are the first to include
both the order-$\alpha_s$ correction to the light-cone distributions
amplitude and the resummation of logarithms of $m_Z^2/m_Q^2$ and are the
first calculations for the $\Upsilon(2S)$ and $\Upsilon(3S)$ final
states. The resummations of logarithms of $m_Z^2/m_Q^2$ that are
associated with the order-$\alpha_s$ and order-$v^2$ corrections are
carried out by making use of the Abel-Pad\'e method. We confirm the
analytic result for the order-$v^2$ correction that was presented in a
previous publication, and we correct the relative sign of the direct and
indirect amplitudes and some choices of scales in that publication. Our
branching fractions for $Z\to J/\psi+\gamma$ and $Z\to
\Upsilon(1S)+\gamma$ differ by $2.0\,\sigma$ and $-4.0\,\sigma$,
respectively, from the branching fractions that are given in the most
recent publication on this topic (in units of the uncertainties that are
given in that publication). However, we argue that the uncertainties in
the rates are underestimated in that publication.

\end{abstract}
\maketitle

\section{Introduction}

The rare decays of the Higgs boson ($H$) to a vector quarkonium ($V$)
and a photon ($\gamma$) have been proposed as processes with which to
measure the Higgs-boson couplings to the charm and bottom quarks
\cite{Bodwin:2013gca}. Even at a high-luminosity LHC, observations of
these decay processes would be challenging. It has been pointed out in
Refs.~\cite{Huang:2014cxa,Grossmann:2015lea} that the decays of the $Z$
boson $Z\to V+\gamma$ could provide means to calibrate the experimental
techniques that might be used to measure the $H\to V+\gamma$ decay rates.

As was emphasized in Ref.~\cite{Bodwin:2013gca}, in the decays $H\to
V+\gamma$, two processes give important contributions to the amplitude:
(1) a {\it direct process}, in which the Higgs boson decays to a
heavy-quark-antiquark pair ($Q\bar{Q}$), which emits a photon and
evolves into a quarkonium; (2) an {\it indirect process}, in which the
Higgs boson decays through a virtual heavy-quark or $W$-boson loop into
a photon and a virtual photon, with the virtual photon decaying into a
heavy quarkonium. In the case of the decays $H\to V+\gamma$, the
indirect process is enhanced for massive particles in the virtual loop
because the Higgs-boson coupling to the loop particle is proportional to
the mass of the particle.

In a classic paper \cite{Guberina:1980dc}, analytic expressions were
given for the direct amplitudes and the corresponding decay rates for
$Z$-boson decays to a photon plus an $S$-wave or a $P$-wave quarkonium.
These expressions were calculated at leading order (LO) in $\alpha_s$,
the QCD running coupling, and at LO in $v^2$, where $v$ is the velocity
of the heavy quark ($Q$) or the heavy antiquark ($\bar Q$) in the
heavy-quarkonium rest frame.

Calculations of exclusive quarkonium production processes can be
simplified by making use of the light-cone approach
\cite{Lepage:1980fj,Chernyak:1983ej}, which yields a systematic
expansion in powers of $m_V/m_{\rm hard}$, where $m_V$ is the quarkonium
mass and $m_{\rm hard}$ is the hard-scattering scale, which is of
order the $Z$-boson mass $m_Z$ in the present case. In the light-cone
approach, nonperturbative effects in the quarkonium system are
parametrized in terms of the quarkonium light-cone distribution
amplitudes (LCDAs). A heavy-quarkonium LCDA can, by virtue of
nonrelativistic QCD (NRQCD) factorization \cite{Bodwin:1994jh}, be
written as a sum of products of short-distance coefficients times NRQCD
long-distance matrix elements (LDMEs) \cite{Jia:2008ep}.

In Ref.~\cite{Luchinsky:2017jab}, calculations of the rates for 
$Z$-boson decays to a photon plus $\eta_c$, $J/\psi$, $\chi_{c0}$, 
$\chi_{c1}$, $\chi_{c2}$, or $h_c$ were presented. These calculations 
were based on the direct amplitude and were carried out at LO in 
$\alpha_s$ and $v^2$ in both the NRQCD and light-cone formalisms.

In Ref.~\cite{Huang:2014cxa}, the decay rates for the processes $Z\to
V+\gamma$, where $V$ is the $J/\psi$ or the $\Upsilon(1S)$, were
computed in the leading-power light-cone approximation, which is valid
up to corrections of order $m_V^2/m_Z^2$. The calculations were carried
out at next-to-leading order (NLO) in $\alpha_s$ and $v^2$.
Reference~\cite{Huang:2014cxa} also gave the first result for the
short-distance coefficient of the order-$v^2$ (relativistic)
corrections. The calculations in Ref.~\cite{Huang:2014cxa} included
contributions from the indirect production process. These contributions
were found to be small, producing effects of less than 1\,\% on the
rates, because, in contrast with the Higgs-boson indirect amplitude, the
$Z$-boson indirect amplitude is not proportional to the mass of the loop
particle. The calculation in Ref.~\cite{Huang:2014cxa} did not include
the effects of resummation of large logarithms of the ratio
$m_Z^2/m_Q^2$, where $m_Q$ is the heavy-quark mass. This resummation was
estimated in Ref.~\cite{Huang:2014cxa} to produce a $1.5\,\%$ effect in
the rate for $Z\to J/\psi +\gamma$.

In Ref.~\cite{Grossmann:2015lea}, the decay rates for the processes
$Z\to V+\gamma$, where $V$ is the $J/\psi$ or the $\Upsilon(1S)$, were
also computed in the leading-power light-cone approximation at NLO in
$v^2$ and at LO in $\alpha_s$.  Logarithms of
$m_Z^2/m_Q^2$  were resummed to all orders in $\alpha_s$ at leading
logarithmic (LL) accuracy. In the case of the order-$v^2$
correction, the resummation of logarithms of $m_Z^2/m_Q^2$ was
carried out by introducing a model for the LCDA whose second moment in
the light-cone momentum fraction $x$ (in the narrow-width
approximation) matches the second $x$ moment of the order-$v^2$ term
in the nonrelativistic expansion of the LCDA. It was found in
Ref.~\cite{Grossmann:2015lea} that the resummation effects are much
larger than the $1.5\,\%$ estimate that was given in
Ref.~\cite{Huang:2014cxa}.

In principle, one can carry out the resummation of logarithms of
$m_Z^2/m_Q^2$ for the order-$v^2$ and order-$\alpha_s$ corrections to the
LCDA  directly, avoiding the unknown uncertainties that are associated
with the introduction of a model LCDA. However, as was pointed out in
Refs.~\cite{Bodwin:2014bpa,Bodwin:2016edd}, the standard approach for
such calculations, namely, expansion in a series in the LO evolution
eigenvectors (Gegenbauer polynomials), fails because the eigenvector
series diverges, even though the evolved LCDA itself is well defined.
This divergence can be traced to the fact that the order-$v^2$ and
order-$\alpha_s$ corrections to the LCDA contain distributions
(generalized functions) \cite{Bodwin:2016edd}. A general solution to this
problem was given in Ref.~\cite{Bodwin:2016edd}, where it was shown that
the evolved LCDA can be obtained by using the so-called Abel-Pad\'e
method to sum the divergent eigenvector series. The Abel-Pad\'e method
allows one to compute the resummation of logarithms of $m_Z^2/m_Q^2$ for
the order-$v^2$ and order-$\alpha_s$ corrections to the LCDA from first
principles.

In the present paper, we compute the decay rates for the processes $Z\to
V+\gamma$, where $V$ is one of the four states $J/\psi$ and
$\Upsilon(nS)$, with $n=1,$ $2$, or $3$. Our computation is carried out
at leading power in the light-cone formalism and through orders
$\alpha_s$ and $v^2$. Logarithms of $m_Z^2/m_Q^2$ are resummed in the
direct amplitude at next-to-leading-logarithmic (NLL) accuracy. The
computations of the rates for $Z\to V+\gamma$ in this paper are the
first to include both the order-$\alpha_s$ corrections to the
LCDA and the resummation of logarithms of $m_Z^2/m_Q^2$. The
calculation includes the effects of the indirect process, as well
as the effects of the direct process.

In comparison with the central values in Ref.~\cite{Huang:2014cxa}, our
branching fraction for $Z\to J/\psi+\gamma$ is shifted by about
$-10\,\%$, which is $-0.5\,\sigma$ in the uncertainties of
Ref.~\cite{Huang:2014cxa}, and our branching fraction for $Z\to
\Upsilon(1S)+\gamma$ is shifted by about $-3\,\%$, which is
$-0.3\,\sigma$ in the uncertainties of Ref.~\cite{Huang:2014cxa}.

In comparison with the central values in Ref.~\cite{Grossmann:2015lea},
our branching fraction for $Z\to J/\psi+\gamma$ is shifted by about
$+12\,\%$, which is $+2.0\,\sigma$ in the uncertainties of
Ref.~\cite{Grossmann:2015lea}, and our branching fraction for $Z\to
\Upsilon(1S)+\gamma$ is shifted by about $-11\,\%$, which is
$-4.0\,\sigma$ in the uncertainties of Ref.~\cite{Grossmann:2015lea}. We
argue that the uncertainties in the rates are underestimated in
Ref.~\cite{Grossmann:2015lea}.

We have also confirmed the result in Ref.~\cite{Huang:2014cxa} for the
short-distance coefficient of the order-$v^2$ correction. Our result for
the relative sign between the direct and indirect  amplitudes differs
from that in Ref.~\cite{Huang:2014cxa}, resulting in positive (negative)
interference for the $J/\psi+\gamma$ [$\Upsilon(nS)+\gamma$] final state.
As the indirect amplitude is small relative to the direct amplitude, the
effect of this sign change is much less than the uncertainties in the
calculation. We have also corrected some choices of scales in the
calculation in Ref.~\cite{Huang:2014cxa}. The effects of these
corrections tend to cancel the effects of the resummations of logarithms
of $m_Z^2/m_Q^2$, which are not included in Ref.~\cite{Huang:2014cxa}.

The remainder of this paper is organized as follows. In
Sec.~\ref{sec:direct} we give the expression for the direct amplitude,
and in Sec.~\ref{sec:resummation} we discuss the resummation of
logarithms of $m_Z^2/m_Q^2$ in the direct amplitude. In
Sec.~\ref{sec:indirect} we give the expression for the indirect
amplitude. Section~\ref{sec:computation} contains a discussion of the
numerical calculation of the rates and the uncertainties in that
calculation. In Sec.~\ref{sec:results}, we present our numerical
results and compare them with results
from previous computations. Finally, in Sec.~\ref{sec:summary}, we
summarize and discuss our results.

\section{Light-cone amplitude for the direct process \label{sec:direct}}
The light-cone amplitude for the direct process of $Z\to V+\gamma$ is
given, up to corrections of relative order $m_V^2/m_Z^2$, by 
\begin{subequations}\label{direct-amp}%
\begin{equation}
i\mathcal{M}_\textrm{dir}^\textrm{LC}(Z\to V+\gamma)
=i\mathcal{A}_\textrm{dir}
\epsilon_{\xi\mu\nu\rho}
\epsilon_{Z}^\xi\epsilon_\gamma^{*\mu}\epsilon_V^{*\nu}
p_\gamma^\rho,
\end{equation}
where
\begin{equation}
i\mathcal{A}_\textrm{dir}
=
-\frac{e e_Q g_Z g_A^Q m_V}{m_Z^2}
f_V^\parallel
\int_{0}^1 dx\, T_H(x,\mu)\phi_V^\parallel(x,\mu).
\end{equation}
\end{subequations}%
Here, $e(>0)$ is the electric charge at momentum scale zero,
$e_Q$ is the fractional charge of the heavy quark $Q$, $f_V^\parallel$
is the decay constant of the longitudinally polarized vector quarkonium
$V$, $\epsilon_Z$ is the $Z$-boson polarization, $\epsilon_V$ is the
quarkonium polarization, $\epsilon_\gamma$ and $p_\gamma$ are the photon
polarization and momentum, respectively, $\mu$ is the renormalization
scale, $x$ is the $Q$ momentum fraction of $V$, which runs from 0
to 1, and $g_Z$ and $g_A$ are defined by
\begin{eqnarray}
g_Z&=&2(\sqrt{2}G_F)^{1/2}m_Z,
\nonumber \\
g_A^Q&=&\frac{1}{2}(T_3^Q)_L.
\end{eqnarray}
Here, $G_F$ is the Fermi constant, and $(T_3^f)_L$ is the eigenvalue
of the weak isospin of the left-handed fermion $f$, whose value is
$+1/2$ for $f=u$, $c$, $t$, $\nu_e$, $\nu_\mu$, $\nu_\tau$, and $-1/2$
for $f=d$, $s$, $b$, $e$, $\mu$, $\tau$. We use the convention
$\epsilon_{0123} = -1$.

The longitudinally  polarized LCDA $\phi_V^\parallel$ is defined in
Refs.~\cite{Ball:1996tb,Ball:1998sk} as
\begin{equation}
\langle V(p)|\bar{Q}(z)\gamma^\mu [z,0] Q(0) |0\rangle
=
-
p^\mu \frac{\epsilon_V^*\cdot z}{p\cdot z}
f_V^\parallel
m_V
\int_0^1 dx\, 
e^{ip\cdot zx}\phi_V^\parallel(x,\mu),
\label{LCDA-defn}
\end{equation}
where $p$ is the quarkonium momentum, $z$ lies along the plus light-cone
direction, and 
\begin{equation}
[z,0]=P\exp\left[
ig_s\int_0^z dx A_a^+(x)T^a
\right]
\end{equation}
is a gauge link that makes the nonlocal operator gauge invariant. Here,
$g_s=\sqrt{4\pi \alpha_s}$, $A^\mu_a$ is the gluon field with the color
index $a=1,$ 2, $\dots$, $N_c^2-1$, $N_c=3$, $T^a$ is a generator of
color SU(3) in the fundamental representation, and the symbol $P$
denotes path ordering. Note that we have included a factor $(-1)$ in
the definition (\ref{LCDA-defn}) relative to the definition in
Refs.~\cite{Ball:1996tb,Ball:1998sk} in order to obtain a positive value
for the decay constant. We note that the definition (\ref{LCDA-defn}) is
equivalent to the definition  of $\phi_V^\parallel$ in
Ref.~\cite{Wang:2013ywc}, from which we take the order-$\alpha_s$
corrections to $\phi_V^\parallel$.

Setting $z$ to $0$ and imposing the normalization condition 
\begin{equation}
\int_0^1 dx\,\phi_V^\parallel(x,\mu)=1,
\end{equation}
we obtain
\begin{equation}
\label{eq:decay-constant-def}
\langle V|\bar{Q}(0)\gamma^\mu Q(0) |0\rangle
=
-
f_V^\parallel
m_V \epsilon_V^{*\mu},
\end{equation}
which allows one to relate the decay constant $f_V^\parallel$ to the
quarkonium electromagnetic decay width $\Gamma(V\to e^+e^-)$:
\begin{equation}
\Gamma(V\to e^+e^-)=\frac{4\pi}{3m_V}\alpha^2(m_V)e_Q^2
f_V^{\parallel 2}.
\label{EM-width}
\end{equation}
Here, $\alpha(m_V)$ is the running electromagnetic coupling at the scale 
$m_V$.

We expand the LCDA at the low-energy scale $\mu_0$, which is of order
$m_Q$, through order $v^2$ and through order $\alpha_s$: 
\begin{equation}
\label{decay-const-expn}
\phi_V^\parallel(x,\mu_0)
=\phi_V^{\parallel(0)}(x,\mu_0)
+\langle v^2\rangle_V
\phi_V^{\parallel(v^2)}(x,\mu_0)
+
\frac{\alpha_s(\mu_0)}{4\pi} \phi_V^{\parallel(1)}(x,\mu_0)
+O(\alpha_s^2,\alpha_s v^2,v^4).
\end{equation}
The quantity $\langle v^2\rangle_V$ is proportional to the
ratio of the NRQCD LDME of order $v^2$ to the NRQCD LDME of
order $v^0$. The general expression for the ratio of the NRQCD LDME 
of order $v^{2k}$ ($k$ a nonnegative integer) to the NRQCD LDME of
order $v^0$ is
\begin{equation}
\langle v^{2k}\rangle_V
=\frac{1}{m_Q^{2k}}
\frac{
\langle V(\bm{\epsilon}_V)|
\psi^\dagger 
(-\tfrac{i}{2}\stackrel{\leftrightarrow}{\bm{\nabla}})^{2k}
\bm{\sigma}\cdot \bm{\epsilon}_V \chi|0\rangle
}{\langle V(\bm{\epsilon}_V)|
\psi^\dagger \bm{\sigma}\cdot \bm{\epsilon}_V \chi|0\rangle}.
\label{LCDA-expn}
\end{equation}
Here, $\psi$ is the two-component (Pauli) spinor field that annihilates
a heavy quark, $\chi^\dagger$ is the two-component spinor field that
annihilates a heavy antiquark, $\bm{\sigma}$ is a Pauli matrix,
$|V(\bm{\epsilon}_V)\rangle$ denotes the vector quarkonium state in the
quarkonium rest frame with spatial polarization $\bm{\epsilon}_V$, and
$m_Q$ denotes the quark pole mass. The light-cone functions on
the right side of Eq.~(\ref{decay-const-expn}) are given by
\begin{subequations}
\begin{eqnarray}
\phi_V^{\parallel(0)}(x,\mu_0)
&=&
\delta(x-\tfrac{1}{2}),
\\
\phi_V^{\parallel(v^2)}(x,\mu_0)
&=&
\frac{\delta^{(2)}(x-\frac{1}{2})}{24},\label{phi-vsq}
\\
\phi_V^{\parallel(1)}(x,\mu_0)
&=&
C_F\theta(1-2x)
\Bigg\{
\left[
\left(4x+\frac{8x}{1-2x}\right)
\left(\log\frac{\mu_0^2}{\overline{m}_Q^2(1-2x)^2}-1\right)
\right]_+
\nonumber \\
&&\quad\quad\quad\quad\quad\quad
+
\left[
\frac{16x(1-x)}{(1-2x)^2}
\right]_{++}
-\left[8x\right]_+
\Bigg\}
+(x\leftrightarrow 1-x).
\label{eq:phiV-alphas}
\end{eqnarray}
\end{subequations}
Here, the $+$ and $++$ functions are defined by
\begin{subequations}
\begin{eqnarray}
\label{def:+and++functions}
\int_0^1 dx\,[f(x)]_{+}g(x)&=&\int_0^1 dx\, 
f(x)[g(x)-g(\tfrac{1}{2})], \\
\int_0^1 dx\,[f(x)]_{++}g(x)&=&\int_0^1 dx\, 
f(x)[g(x)-g(\tfrac{1}{2})-g'(\tfrac{1}{2})(x-\tfrac{1}{2})].
\end{eqnarray}
\end{subequations}
The order-$\alpha_s$ light-cone function
$\phi_V^{\parallel(1)}(x,\mu_0)$ was computed in
Ref.~\cite{Wang:2013ywc}. In Eq.~(\ref{eq:phiV-alphas}), we have
replaced the pole mass $m_Q$ with $\overline{m}_Q$, the
modified-minimal-subtraction ($\overline{\rm MS}$) mass, since the pole
mass is ill defined, owing to renormalon ambiguities.  This change
affects the expression for $\phi_V^\parallel$ at relative order
$\alpha_s^2$. The order-$v^2$ light-cone function
$\phi_V^{\parallel(v^2)}(x,\mu_0)$ was computed in
Ref.~\cite{Huang:2014cxa}. It can be inferred from the computation for
the quarkonium transverse LCDA in Ref.~\cite{Bodwin:2014bpa} by using 
the fact that the relativistic corrections to the LCDA are independent of
the quarkonium spin \cite{Braguta:2007fh}. It can also be inferred from
the calculation in Ref.~\cite{Wang:2017bgv} for $S$-wave $B_c$ mesons in
the limit $m_c=m_b$. We have verified this result by using the NRQCD
formalism to compute the complete order-$v^2$ contribution to the direct
amplitude, which includes the order-$v^2$ contribution to
$\phi_V^\parallel$ in Eq.~(\ref{phi-vsq}) and the order-$v^2$
contribution to the decay constant $f_V^\parallel$, and making use of
the known order-$v^2$ contribution to $f_V^\parallel$ [see
Eq.~(\ref{fV-expn}) below].

The decay constant $f_V^\parallel$ is given by 
\begin{equation}
f_V^\parallel
=
\frac{\sqrt{2N_c}\sqrt{2m_V}}{m_V}\Psi_V(0)
\left[1-\frac{1}{6}\langle v^2\rangle_V
-8\frac{\alpha_s(\mu_0) C_F}{4\pi}
+O(\alpha_s^2,\alpha_s v^2,v^4)
\right],
\label{fV-expn}
\end{equation}
where $C_F=(N_c^2-1)/(2N_c)$ and $C_A=N_c=3$ for color $\textrm{SU}(3)$.
We note that $f_V^\parallel$, as defined in Eq.~(\ref{EM-width}), is
scale invariant. Hence, the dependence of the expression in brackets on
the right side of Eq.~(\ref{fV-expn}) on the scale $\mu_0$ implies that
$\Psi_V(0)$ depends on $\mu_0$ in such a way as to render the complete
expression scale invariant. The quarkonium wave function at the origin
$\Psi_V(0)$ is related to the LO NRQCD LDME
\cite{Bodwin:1994jh}:
\begin{equation}
\Psi_V(0)=\frac{1}{\sqrt{2N_c}}\langle V(\bm{\epsilon}_V)| 
\psi^\dagger \bm{\sigma}\cdot \bm{\epsilon}_V \chi|0\rangle.
\end{equation}
The LO and order-$\alpha_s$ contributions to $f_V^\parallel$ were
computed in Ref.~\cite{Wang:2013ywc}. The order-$v^2$ contribution to
$f_V^\parallel$ was computed in Ref.~\cite{Huang:2014cxa}. It can be
inferred from the  order-$v^2$ contribution to the quarkonium
electromagnetic decay rate in Ref.~\cite{Bodwin:1994jh}.

In this paper, we will use Eq.~(\ref{EM-width}) to compute
$f_V^\parallel$ from the measured electromagnetic decay widths. As was
pointed out in  Ref.~\cite{Grossmann:2015lea}, this procedure eliminates
the uncertainties in the calculation that arise from the use of
Eq.~(\ref{fV-expn}) in conjunction with phenomenological determinations
of $\Psi_V(0)$ and $\langle v^2\rangle_V$. Equation~(\ref{fV-expn}) was
used in the calculation in Ref.~\cite{Huang:2014cxa}. We defer a
discussion of the impact of that choice to Sec.~\ref{sec:results}.

The hard-scattering kernel
$T_H(x,\mu)$ for the process $Z\to V+\gamma$, through order 
$\alpha_s$, is \cite{Wang:2013ywc}
\begin{subequations}
\begin{equation}
T_H(x,\mu)= 
T_H^{(0)}(x,\mu)
+\frac{\alpha_s(\mu)}{4\pi} 
T_H^{(1)}(x,\mu),
\end{equation}
where
\begin{eqnarray}
\label{eq:TH-x}
T_H^{(0)}(x,\mu)
&=&
\frac{1}{x(1-x)},
\\
T_H^{(1)}(x,\mu)
&=&
C_F\frac{1}{x(1-x)}
\bigg\{
\big[3+2x\log(1-x)+2(1-x)\log x\big]
\left(\log\frac{m_Z^2}{\mu^2}-i\pi\right)
\nonumber \\
&&\quad\quad
+x\log^2(1-x)+(1-x)\log^2 x
-(1-x)\log(1-x)-x\log x-9
\bigg\}.\phantom{xxx}
\end{eqnarray}
\end{subequations}

\section{Resummation of logarithms in the direct amplitude 
\label{sec:resummation}}
\subsection{The Gegenbauer expansion of the amplitude}

The scale evolution of the LCDA can be computed conveniently by
expanding the LCDA in Gegenbauer polynomials, which are the eigenvectors
of the LO evolution kernel. The Gegenbauer expansion of the LCDA is
\begin{equation}
\phi_V^\parallel(x,\mu)\equiv 
\sum_{n=0}^\infty
\phi_n^\parallel(\mu)\,x(1-x)\,C_n^{(3/2)}(2x-1),
\end{equation}
where $\phi_n^\parallel$ is the $n$th Gegenbauer moment of
$\phi_V^\parallel$, which can be found by making use of the
orthogonality property of the Gegenbauer polynomials:
\begin{equation}
\phi_n^\parallel(\mu)=N_n
\int_0^1 dx\,C_n^{(3/2)}(2x-1) \phi_V^\parallel(x,\mu),
\end{equation} 
where
\begin{equation}
N_n=\frac{4(2n+3)}{(n+1)(n+2)}.
\end{equation}
Note that $\phi_n^\parallel(\mu)$ vanishes for odd $n$ because
$\phi_V^\parallel(x,\mu)$ is symmetric about $x=1/2$. We define the
LO, order-$v^2$, and order-$\alpha_s$ Gegenbauer moments of
$\phi_V^\parallel$ as follows:
\begin{equation}
\phi_n^\parallel(\mu)
\equiv
\phi_n^{\parallel(0)}(\mu)
+\langle v^2\rangle_V
\phi_n^{\parallel(v^2)}(\mu)
+\frac{\alpha_s(\mu_0)}{4\pi}
\phi_n^{\parallel(1)}(\mu)+O(\alpha_s^2,\alpha_s v^2,v^4).
\end{equation}

The moments $\phi_n^\parallel(\mu)$ can be written in terms of
the moments $\phi_n^\parallel(\mu_0)$ and an evolution matrix
$U_{nk}(\mu,\mu_0)$:
\begin{equation}
\phi_n^\parallel(\mu) =
\sum_{k=0}^n U_{nk}(\mu,\mu_0)\phi_k^\parallel(\mu_0).
\end{equation}
The LL and NLL expressions for $U_{nk}(\mu,\mu_0)$ can be found in the
Appendix. 

The Gegenbauer expansion of the hard-scattering kernel is given by
\begin{equation}
T_H(x,\mu)=\sum_{n=0}^\infty N_n T_n(\mu) C_n^{(3/2)}(2x-1),
\end{equation}
where $T_n$ is the $n$th Gegenbauer moment of $T_H$, which can be
found by making use of the orthogonality property of the Gegenbauer
polynomials:
\begin{equation}
T_n(\mu)=\int_0^1 dx\, x(1-x)C_{n}^{(3/2)}(2x-1) T_H(x,\mu).
\end{equation}
We define the LO and order-$\alpha_s$ contributions to $T_n$ as follows:
\begin{equation}
T_n(\mu)=T_n^{(0)}(\mu)+\frac{\alpha_s(\mu)}{4\pi}T_n^{(1)}(\mu)+O(\alpha_s^2).
\end{equation}

Making use of the orthogonality property of the Gegenbauer   
polynomials again, we can write the light-cone amplitude as
\begin{equation}
{\cal M}(\mu)=\int_0^1 dx\, T_H(x,\mu)\phi_V^\parallel(x,\mu)
=\sum_{n=0}^\infty T_n(\mu) \phi_n^\parallel(\mu).
\end{equation}
We also define a decomposition of ${\cal M}$ into the LO, order-$v^2$, and 
order-$\alpha_s$ contributions:
\begin{eqnarray}\label{Mdir}
\mathcal{M}(\mu)
&=&
\mathcal{M}^{(0,0)}(\mu)
+\langle v^2\rangle_{V}
\mathcal{M}^{(0,v^2)}(\mu)
+\frac{\alpha_s(\mu)}{4\pi}
\mathcal{M}^{(1,0)}(\mu)
+\frac{\alpha_s(\mu_0)}{4\pi}
\mathcal{M}^{(0,1)}(\mu)\nonumber\\
&&+O(\alpha_s^2,\alpha_s v^2,v^4),
\end{eqnarray}
where
\begin{equation}
\mathcal{M}^{(i,j)}(\mu)
=\sum_{n=0}^\infty T_{n}^{(i)}(\mu)
\phi_{n}^{\parallel(j)}(\mu).
\label{amplitude-series}
\end{equation}

By choosing the scale $\mu$ in ${\cal M}(\mu)$ to be $m_Z$, we guarantee
that $T_n(\mu)$ contains no large logarithms. We choose the initial
scale of the LCDAs to be $\mu_0=\overline{m}_Q$. Then, logarithms of
$m_Z^2/\overline{m}_Q^2$ are resummed by the evolution of
$\phi_n^\parallel$ from the scale $\mu_0=\overline{m}_Q$ to the
scale $\mu=m_Z$.

Using Eq.~(\ref{Mdir}), we find that the resummed direct amplitude 
is given by 
\begin{eqnarray}\label{Adir}%
i\mathcal{A}_\textrm{dir}^\textrm{LC}
=
-\frac{e e_Q g_Z g_A^Q m_V}
{m_Z^2} f_V^\parallel
\Big[&&\!\!
\mathcal{M}^{(0,0)}(\mu)
+\frac{\alpha_s(\mu)}{4\pi}\mathcal{M}^{(1,0)}(\mu)
+\frac{\alpha_s(\mu_0)}{4\pi}\mathcal{M}^{(0,1)}(\mu)
\nonumber \\
&&\!\!+\langle v^2 \rangle_V \mathcal{M}^{(0,v^2)}(\mu)
\Big]+O(\alpha_s^2,\alpha_s v^2,v^4).
\end{eqnarray}
We use this expression in our numerical calculations. We make use 
of expressions for the evolution through NLL accuracy, which are given in 
the Appendix.

\subsection{The Abel-Pad\'e method}

The sum over $n$ in Eq.~(\ref{amplitude-series}) diverges for 
$\mathcal{M}^{(0,v^2)}$ and $\mathcal{M}^{(0,1)}$ 
\cite{Bodwin:2014bpa,Bodwin:2016edd}. As was explained in 
Ref.~\cite{Bodwin:2016edd}, such divergences can arise because the 
light-cone distributions contain generalized functions (distributions), 
rather than ordinary functions. In Ref.~\cite{Bodwin:2016edd}, it was
shown that one can define the generalized functions as a limit of
ordinary functions, which leads one to compute $\mathcal{M}^{(i,j)}$
as follows:
\begin{equation}
{\cal M}^{(i,j)}(\mu)=
\lim_{z\to 1} \sum_{n,m=0}^\infty T_m^{(i)}(\mu) U_{mn}(\mu,\mu_0) 
z^n \phi_{n}^{\parallel(j)}(\mu_0).
\label{amplitude-series-abel}
\end{equation}
The expression in Eq.~(\ref{amplitude-series-abel}) is the Abel
summation of the eigenfunction series for $\phi_V^\parallel(x,\mu_0)$.
In  Ref.~\cite{Bodwin:2016edd}, the Abel summation was erroneously
applied to $\phi_V^\parallel(x,\mu)$. (See Ref.~\cite{Bodwin:2017wdu}.)
We have corrected that error here.
The correction amounts to the replacement of $z^m$ with $z^n$ in
Eq.~(\ref{amplitude-series-abel}).

One can improve upon the convergence of the series in
Eq.~(\ref{amplitude-series-abel}) in the limit $z\to 1$, by constructing
a Pad\'e approximant for the $n$th partial sum before taking the limit
$z\to 1$. The use of the Pad\'e approximant is effective in improving
the convergence of the series because it provides an approximate
analytic continuation for the function of $z$ that is represented by the
series. That analytic continuation is valid beyond the radius of
convergence of the series, which is typically $|z|=1$. The Abel-Pad\'e
method was tested extensively against known analytic results for
$\mathcal{M}^{(i,j)}$ in Ref.~\cite{Bodwin:2016edd}, and it converged
rapidly to the correct value in all cases. We will use it throughout
this paper to evaluate $\mathcal{M}^{(i,j)}$.

\section{Amplitude for the indirect process\label{sec:indirect}}

The amplitude for the indirect decay amplitude contains the
axial-vector-vector triangle diagram as a subdiagram. The amplitude
for the axial-vector-vector triangle diagram is given in
Ref.~\cite{Hagiwara:1990dx}. In that paper, the conventions for
$\gamma_5$ and the completely antisymmetric tensor
$\epsilon_{\xi\mu\nu\rho}$ are not specified. We fix the overall sign of
the triangle amplitude in Ref.~\cite{Hagiwara:1990dx} in our conventions
by requiring that it give the correct axial-vector anomaly. Then, we
find that the indirect amplitude for the decay of a $Z$ boson to a
photon plus a virtual photon is given by
\begin{equation}
i\mathcal{M}(Z\to \gamma\gamma^*)
=
-g_Ze^2 \sum_f e_f^2g_A^f f_2^f m_V^2
\epsilon^\xi\epsilon^{*\mu}\epsilon^{*\nu}
\epsilon_{\xi\mu\nu\rho} p_\gamma^\rho,
\label{Zgamgamstar}
\end{equation}
where $f$ denotes any fermion that can appear in the loop in
the triangle diagram and 
\begin{eqnarray}
\label{eq:def-fi}
f_2^f
=
\frac{1}{\pi^2}
\int_0^1\! dz_1 \int_0^1\! dz_2 
\int_0^1\! dz_3 \,\delta(1-z_1-z_2-z_3)
\frac{z_2z_3}{m_f^2-z_1z_2p_\gamma^2-z_2z_3 m_V^2-z_3z_1 m_Z^2}.
\end{eqnarray}
Here, $\epsilon^\xi$, $\epsilon^{*\mu}$, and $\epsilon^{*\nu}$ are the 
polarizations of the $Z$ boson, real photon, and virtual photon, 
respectively, and $p_\gamma$ is the momentum of the real photon 
($p_\gamma^2=0$).\footnote{Owing to the masslessness of the photon
and the orthogonality the $Z$-boson momentum and polarizations, some
terms that appear in the complete triangle-diagram amplitude and
that contribute to the axial-vector anomaly do not contribute to
Eq.~(\ref{Zgamgamstar}).} Then, following Refs.~\cite{Bodwin:2014bpa,
Bodwin:2013gca}, we obtain the indirect amplitude for process $Z\to
V+\gamma$:
\begin{subequations}\label{indirect-amp}
\begin{eqnarray}
i\mathcal{M}_\textrm{ind}
(Z\to V+\gamma)
&=&
i\mathcal{M}(Z\to \gamma\gamma^*)
\frac{-i}{m_V^2}
(-ieg_{V\gamma})
\nonumber \\
&=&
i\mathcal{A}_\textrm{ind}
\epsilon_{\xi\mu\nu\rho}
\epsilon_{Z}^\xi \epsilon_\gamma^{*\mu}\epsilon_V^{*\nu}
p_\gamma^\rho,
\end{eqnarray}
where
\begin{equation}
i\mathcal{A}_\textrm{ind}=
g_Zg_{V\gamma}\big[\sqrt{4\pi\alpha(m_V)}\big]^2
\sqrt{4\pi\alpha(0)}
\sum_{f}
e_f^2g_A^f f_2^f.
\label{Aind}
\end{equation}
\end{subequations}%
Here, $g_{V\gamma}$ is given by
\begin{equation}
g_{V\gamma}
=
-\frac{e_Q}{|e_Q|}
\left[
\frac{3m_V^3\Gamma(V\to e^+ e^-)}{4\pi\alpha^2(m_V)}
\right]^{1/2}.
\end{equation}

The relative sign between the direct amplitude in Eq.~(\ref{direct-amp})
and the indirect amplitude in Eq.~(\ref{indirect-amp}) disagrees with
the relative sign that was found in Ref.~\cite{Huang:2014cxa}. That is,
we find that the direct and indirect amplitudes interfere constructively
for the process $Z\to J/\psi +\gamma$ and interfere destructively for
the processes $Z\to \Upsilon(nS) +\gamma$.

\section{Computation of the decay rates \label{sec:computation}}
\subsection{Decay rate}

The rate for the decay of a $Z$ boson into a vector quarkonium plus a
photon is easily seen to be
\begin{eqnarray}
\Gamma(Z\to V+\gamma)&=&\frac{1}{48\pi m_Z}\sum_{\rm pol}
|\mathcal{M}_{\rm dir}(Z\to V+\gamma)+\mathcal{M}_{\rm ind}(Z\to
V+\gamma)|^2\nonumber\\
&=&\frac{m_Z^3}{96\pi m_V^2}|\mathcal{A}_{\rm dir}
+\mathcal{A}_{\rm ind}|^2,
\end{eqnarray}
where $\mathcal{A}_{\rm dir}$ is given in Eq.~(\ref{Adir}),
$\mathcal{A}_{\rm ind}$ is given in  Eq.~(\ref{Aind}), and we have
dropped terms of order $m_V^2/m_Z^2$. In evaluating the expression for
$\mathcal{A}_{\rm dir}$ in Eq.~(\ref{Adir}), we take the hard-scattering
scale $\mu$ to be $m_Z$, and we take the initial scale $\mu_0$ to be the
heavy-quark $\overline{\rm MS}$ mass $\overline{m}_Q$. The typical
momentum scale of loop corrections to the LCDA and to $f_V^\parallel$ is
the pole mass, and, so, the pole mass would be a natural choice for
$\mu_0$.  However, the pole mass is ill defined, as we have already
mentioned, owing to renormalon ambiguities, and the presence of
pole-mass renormalons could impact the convergence of the perturbation
series unfavorably in higher orders. Therefore, we choose to take
$\mu_0=\overline{m}_Q$. In applying the Abel-Pad\'e method to the
expression for $\mathcal{A}_{\rm dir}$ in Eq.~(\ref{Adir}), we take 100
terms in the eigenfunction expansion and use a $50\times 50$ Pad\'e
approximant. As we have mentioned, in order to minimize uncertainties in
$f_V^\parallel$, we follow Ref.~\cite{Grossmann:2015lea} and compute
$f_V^\parallel$ from the leptonic width of the quarkonium, using
Eq.~(\ref{EM-width}), instead of using the perturbative expression in
Eq.~(\ref{fV-expn}).

\subsection{Numerical inputs}

We take the pole masses to be the one-loop values $m_c=1.483$~GeV and
$m_b=4.580$~GeV, we take the $\overline{\rm MS}$ masses to be
$\overline{m}_c=1.275$~GeV and $\overline{m}_b=4.18$~GeV, and we take
$m_Z=91.1876$~GeV and $\Gamma(Z)=(2.4952\pm0.0023)$~GeV. We also take
$\alpha(m_{J/\psi})=1/132.642$ and $\alpha(m_{\Upsilon(nS)})=1/131.015$.
Our values for $|\Psi_V(0)|^2$, $\langle v^2\rangle_V$, and
$f_V^\parallel$ are shown in Table~\ref{num-inputs}.
\begin{table}[h]
\begin{center}
\begin{tabular}{llll} 
\hline
$\phantom{xx}V\phantom{x}$ & 
$\phantom{x}|\Psi_V(0)|^2~(\textrm{GeV}^3)$&
$\phantom{xxxxxxx}\langle v^2\rangle_V$ &
$\phantom{x}f_V^\parallel~(\textrm{MeV})$
\\
\hline\hline
$\phantom{x}J/\psi$\phantom{x} & 
$\phantom{-}0.0729\pm 0.0109$ & 
$\phantom{xx}\phantom{-}0.201\pm 0.064$ &
$\phantom{x}403.0\pm 5.1$
\\
$\phantom{x}\Upsilon(1S)\phantom{xx}$&
$\phantom{-}0.512\pm 0.035$ &
$\phantom{x}-0.00920\pm 0.0105\phantom{xx}$ &
$\phantom{x}683.8\pm 4.6\phantom{x}$
\\
$\phantom{x}\Upsilon(2S)\phantom{x}$ & 
$\phantom{-}0.271\pm 0.019$ & 
$\phantom{xx}\phantom{-} 0.0905\pm 0.0109$ &
$\phantom{x}475.6\pm 4.3$
\\
$\phantom{x}\Upsilon(3S)\phantom{x}$ & 
$\phantom{-}0.213\pm 0.015$ &
$\phantom{xx}\phantom{-}0.157\pm 0.017$ &
$\phantom{x}411.3\pm3.7$
\\
\hline
\end{tabular}
\caption{\label{num-inputs}
Values of $|\Psi_V(0)|^2$, $\langle v^2\rangle_V$, and $f_V^\parallel$
for $V=J/\psi$ and $\Upsilon(nS)$. The values for $|\Psi_V(0)|^2$ and
$\langle v^2\rangle_V$ have been taken from
Refs.~\cite{Bodwin:2007fz,Chung:2010vz}, except for the uncertainties in
$\langle v^2\rangle_{\Upsilon(1S)}$ and $\langle
v^2\rangle_{\Upsilon(2S)}$, which are described in the text. The values
for $f_V^\parallel$ have been computed by making use of
Eq.~(\ref{EM-width}).}
\end{center}
\end{table}
We do not use the values for $|\Psi_V(0)|^2$ in our calculations, but we
include them here for purposes of later comparison with the calculations
in Ref.~\cite{Huang:2014cxa}. We use the values for $|\Psi_V(0)|^2$ and
$\langle v^2\rangle_V$ from Refs.~\cite{Bodwin:2007fz,Chung:2010vz},
except in the cases of $\langle v^2\rangle_{\Upsilon(1S)}$ and $\langle
v^2\rangle_{\Upsilon(2S)}$. As was explained in
Ref.~\cite{Bodwin:2016edd}, the uncertainties for $\langle
v^2\rangle_{\Upsilon(1S)}$ and $\langle v^2\rangle_{\Upsilon(2S)}$ were
probably underestimated in Ref.~\cite{Chung:2010vz}. We use the larger
uncertainties for these quantities that are given in
Ref.~\cite{Bodwin:2016edd}. 

\subsection{Sources of uncertainties}

In calculating the decay rates, we take into account uncertainties in
both the direct and indirect amplitudes, as is described below. We also 
include the uncertainty in the $Z$-boson total width in computing 
branching fractions. 
We compute the overall uncertainties in the rates by 
making use of the method that is described in Sec.~VIE of 
Ref.~\cite{Bodwin:2016edd}. That is, we find the extrema of the rate 
for values of the input parameters that lie within a hyperellipse that is 
centered at the central values of the input parameters and whose 
semimajor axes have lengths that are equal to the uncertainties in the 
input parameters.  

\subsubsection{Direct amplitude}

In the direct amplitude, we include the uncertainties that arise from
the uncertainties in $f_V^\parallel$ and $\langle v^2\rangle_{V}$. 
We also include the uncertainties that arise from uncalculated
corrections of order $\alpha_s^2$, order $\alpha_s v^2$, and order
$v^4$. We estimate the uncertainties from these uncalculated
corrections, relative to the lowest nontrivial order in the direct
amplitude, to be $\{[C_F
C_A\alpha_s^2(\overline{m}_Q)/\pi^2]^2+[C_F\alpha_s(\overline{m}_Q)
v^2/\pi]^2+[(1/5)v^4]^2\}^{1/2}$ for the real part of the direct
amplitude and
$\{[C_A\alpha_s(\overline{m}_Q)/\pi]^2+[v^2]^2\}^{1/2}$ for the
imaginary part of the direct amplitude. (Note that the real part of the
direct amplitude starts in absolute order $\alpha_s^0$ and the imaginary
part of the direct amplitude starts in absolute order $\alpha_s$.) The
coefficient $1/5$ in the $v^4$ uncertainty in the direct amplitude is
the known short-distance coefficient for the order-$v^4$ correction,
which arises from the expression \cite{Braguta:2007fh} for the $2k$th
$x$ moment of the LCDA $\langle x^{2k}\rangle$ in terms of the
order-$v^{2k}$ LDME ratio $\langle v^{2k}\rangle$ [see
Eq.~(\ref{LCDA-expn})]:
\begin{equation}
\langle x^{2k}\rangle=\frac{\langle v^{2k}\rangle}{2k+1}.
\label{x-moment-LDME}
\end{equation}
We take $v^2=0.3$ for the $J/\psi$ and $v^2=0.1$
for the $\Upsilon(nS)$ states. We also include an uncertainty of
$m_V^2/m_Z^2$ in order to account for uncalculated corrections of order
$m_V^2/m_Z^2$.

\subsubsection{Indirect amplitude}

In indirect amplitude, we include uncertainties that arise from the
uncertainties in the leptonic-decay widths of the quarkonia. We assume
that the uncertainties in the leptonic-decay widths are 2.5\,\% for the
$J/\psi$, 1.3\,\% for the $\Upsilon(1S)$, and 1.8\,\% for the
$\Upsilon(2S)$ and $\Upsilon(3S)$ states. Again, we include an
uncertainty of $m_V^2/m_Z^2$ in order to account for uncalculated
corrections of order $m_V^2/m_Z^2$.

\section{Numerical results and comparisons with previous calculations 
\label{sec:results}}
\subsection{Results}

Our results for the branching fractions of the $Z$ boson into
$J/\psi+\gamma$ and $\Upsilon(nS)+\gamma$ are given in
Table~\ref{tab:num-rates}.
\begin{table}[h]
\begin{center}
\begin{tabular}{llll} 
\hline
$\phantom{xx}V\phantom{x}$ & 
$\phantom{xx}{\rm Br}(Z\to V+\gamma)$ (this work)&
$\phantom{xx}{\rm Br}(Z\to V+\gamma)$ (Ref.~\cite{Huang:2014cxa}) &
$\phantom{xx}{\rm Br}(Z\to V+\gamma)$ (Ref.~\cite{Grossmann:2015lea})
\\
\hline\hline
$\phantom{x}J/\psi$\phantom{x} & 
$\phantom{xx}8.96^{+1.51}_{-1.38} \times 10^{-8}$  &
$\phantom{xx}(9.96\pm1.86) \times 10^{-8} $ &
$\phantom{xx}8.02^{+0.46}_{-0.44} \times 10^{-8} $ 
\\
$\phantom{x}\Upsilon(1S)\phantom{xx}$&
$\phantom{xx}4.80^{+0.26}_{-0.25} \times 10^{-8}$ &
$\phantom{xx}(4.93\pm 0.51)\times 10^{-8}\phantom{xx}$ &
$\phantom{xx}5.39^{+0.17}_{-0.15} \times 10^{-8} $ 
\\
$\phantom{x}\Upsilon(2S)\phantom{x}$ & 
$\phantom{xx}2.44^{+0.14}_{-0.13}\times 10^{-8}$ &
$\phantom{xxxxxxxxx}-$ &
$\phantom{xxxxxxxxx}-$ 
\\
$\phantom{x}\Upsilon(3S)\phantom{x}$ & 
$\phantom{xx}1.88^{+0.11}_{-0.10} \times 10^{-8}$ &
$\phantom{xxxxxxxxx}-$ &
$\phantom{xxxxxxxxx}-$ 
\\
\hline
\end{tabular}
\caption{\label{tab:num-rates}
The branching fractions of $Z\to V+\gamma$
for $V=J/\psi$ and $\Upsilon(nS)$. Our results are shown in the first 
column, and the results from from Refs.~\cite{Huang:2014cxa}
and \cite{Grossmann:2015lea} are shown in the last two columns.
}
\end{center}
\end{table}
For purposes of comparison, we also show the branching fractions from
Refs.~\cite{Huang:2014cxa} and \cite{Grossmann:2015lea}.

As was found in Ref.~\cite{Huang:2014cxa} and noted in
Ref.~\cite{Grossmann:2015lea}, we find that the effect of the indirect
amplitude is small. The inclusion of the indirect amplitude changes the
rate by +1.0\,\% for $Z\to J/\psi+\gamma$, by $-1.1$\,\% for $Z\to
\Upsilon(1S)+\gamma$, by $-1.1$\,\% for $Z\to \Upsilon(2S)+\gamma$, and
by $-1.0$\,\% for $Z\to \Upsilon(3S)+\gamma$.

We also find that the effects of NLL summation are small. The
inclusion of NLL resummation changes the rate by $+2.5\,\%$ for $Z\to
J/\psi+\gamma$, by $+1.9\,\%$ for $Z\to \Upsilon(1S)+\gamma$, by
$1.8\,\%$ for $Z\to \Upsilon(2S)+\gamma$, and by $1.8\,\%$ for
$Z\to \Upsilon(3S)+\gamma$.

Our results for the branching fractions differ considerably from the
results in Refs.~\cite{Huang:2014cxa} and \cite{Grossmann:2015lea}, in
both the central values and in the uncertainties. We now discuss in
detail the reasons for those differences.

\subsection{Comparison with the results from Ref.~[2]}

Our branching fraction for $Z\to J/\psi+\gamma$ differs from that in
Ref.~\cite{Huang:2014cxa} by $-10\,\%$, which is about $-0.5\,\sigma$ 
in the uncertainties of Ref.~\cite{Huang:2014cxa}. Our branching
fraction for $Z\to \Upsilon(1S)+\gamma$ differs from that in
Ref.~\cite{Huang:2014cxa} by $-3\,\%$, which is about $-0.3\,\sigma$ in
the uncertainties of Ref.~\cite{Huang:2014cxa}.

These differences arise from several sources: (1) we have corrected the
value of the scale of $\Psi_V(0)$ that was used in
Ref.~\cite{Huang:2014cxa}; (2) we have corrected the value of the
scale of $\alpha_s$ in the order-$\alpha_s$ corrections to $f_V$ that
was used in Ref.~\cite{Huang:2014cxa}; (3) in the direct amplitude,
we have absorbed the order-$\alpha_s$ and order-$v^2$ NRQCD corrections
to $f_V$ in Eq.~(\ref{fV-expn}) into an overall factor $f_V$ that is
determined from the quarkonium electronic decay width, whereas these
corrections were computed from the NRQCD expansion and incorporated
additively into the direct amplitude in Ref.~\cite{Huang:2014cxa}; (4)
we have found a relative sign between the indirect and direct amplitudes
that is opposite to the sign that was given in
Ref.~\cite{Huang:2014cxa}; (5) we have resummed logarithms of
$m_Z^2/m_Q^2$, which were not resummed in Ref.~\cite{Huang:2014cxa};
(6) we have chosen $\mu_0=\overline{m}_Q$ instead of $\mu_0=m_Q$, and we
have replaced $m_Q$ with $\overline{m}_Q$ in the expression for
$\phi_V^{\parallel(1)}$ in Eq.~(\ref{eq:phiV-alphas}). In
Table~\ref{tab:gross-HP-uncertainty}, the effects on the branching
fractions of the corrections that correspond to these differences are
shown. The fractional change in the branching fraction from each
correction depends on the order in which the corrections are
incorporated into the calculation. In
Table~\ref{tab:gross-HP-uncertainty}, the fractional changes are
computed by incorporating the corrections in the order (1), (2), (3),
(4), (5), (6). For each quarkonium state, the product of
fractional changes gives the fractional change between our result and
that of Ref.~\cite{Huang:2014cxa}. As can be seen from
Table~\ref{tab:gross-HP-uncertainty}, the effects of corrections (1),
(2), (3), and (5) are quite large. However, they tend to cancel each
other, and, consequently, 
our results for branching fractions do not differ so
greatly from those in Ref.~\cite{Huang:2014cxa}. We now discuss the
corrections to the calculation in Ref.~\cite{Huang:2014cxa} in detail.

\begin{table}[h]
\begin{center}
\begin{tabular}{lllllll} 
\hline
$\phantom{x}V\phantom{x}$ & 
\phantom{xx}(1) &
\phantom{xx}(2) &
\phantom{xx}(3) &
\phantom{xx}(4) &
\phantom{xx}(5) &
\phantom{xx}(6)
\\
\hline\hline
$\phantom{x}J/\psi$\phantom{x} & 
$+28.19\,\%$\phantom{xx} &
$-34.73\,\%$\phantom{xx} &
$-12.69\,\%$\phantom{xx} &
$+2.28\,\%$\phantom{xx} &
$+17.54\,\%$\phantom{xx} &
$+2.36\,\%$ \phantom{x}
\\
$\phantom{x}\Upsilon(1S)\phantom{xxx}$&
$+8.13\,\%$ &
$-16.96\,\%$ &
$-2.50\,\%$ &
$-2.40\,\% $ &
$+11.34\,\%$ &
$+1.16\,\%$ 
\\
\hline
\end{tabular}
\caption{\label{tab:gross-HP-uncertainty}
Effects on the branching fractions of corrections to the calculation
in Ref.~\cite{Huang:2014cxa}. The corrections (1)--(6) are described in 
the text.
}
\end{center}
\end{table}

In Ref.~\cite{Huang:2014cxa}, the decay constant $f_V^\parallel$ was
computed by making use of the perturbative expression in
Eq.~(\ref{fV-expn}). As we have mentioned, this results in greater
uncertainties in the calculations. As implemented in
Ref.~\cite{Huang:2014cxa}, it also leads to shifts in the central
values. The reason for this is that the value for $\Psi_V(0)$ that was
used in Ref.~\cite{Huang:2014cxa} was extracted from
Ref.~\cite{Bodwin:2006dn} at the scale $m_V$, while the initial scale
$\mu_0$ in Ref.~\cite{Huang:2014cxa} was taken to be $m_Q$.
Therefore, the value of $\Psi_V(0)$ from Ref.~\cite{Bodwin:2006dn}
should have been corrected as follows in order to account for the change
in the initial scale:
\begin{equation}
|\Psi_V(0)|_{\mu=m_Q}
=
\frac{\displaystyle
1-\frac{\langle v^2\rangle_V}{6}
-8\frac{C_F\alpha_s(m_V)}{4\pi}}
{\displaystyle
1-\frac{\langle v^2\rangle_V}{6}
-8\frac{C_F\alpha_s(m_Q)}{4\pi}}
\,|\Psi_V(0)|_{\mu=m_V}.
\label{scale-corr}
\end{equation}
The fraction on the right side of Eq.~(\ref{scale-corr}) gives
correction~(1), which produces a correction of $+28\,\%$ in the
rate of $Z\to J/\psi+\gamma$ and a correction of $+8\,\%$ in the
rate of $Z\to \Upsilon(1S)+\gamma$.

In the expression for the direct amplitude in Ref.~\cite{Huang:2014cxa},
there are contributions that are proportional to
$-8\alpha_s(m_Z)C_F/(4\pi)-\langle v^2\rangle_{V}/6$. These
contributions arise when one expresses $f_V^\parallel$ in terms of
$\Psi_V(0)$, as in Eq.~(\ref{fV-expn}). However, the argument of
$\alpha_s$ should be $m_Q$, rather than $m_Z$.\footnote{This
incorrect scale choice originated in Eq.~(126) of
Ref.~\cite{Wang:2013ywc} and propagated to Ref.~\cite{Huang:2014cxa}.}
This change of scale accounts for correction~(2), which produces a
correction of $-35\,\%$ in the rate of $Z\to J/\psi+\gamma$ and a
correction of $-17\,\%$ in the rate of $Z\to \Upsilon(1S)+\gamma$.

In the direct amplitude, one can absorb the order-$\alpha_s$ and
order-$v^2$ contributions in the NRQCD expansion of $f_V^\parallel$ in
Eq.~(\ref{fV-expn}) into an overall factor. 
In our calculation, we express the
direct amplitude in terms of the value of $f_V^\parallel$ that one
obtains directly from the electronic width of the quarkonium [see
Eq.~(\ref{direct-amp})]. As we have mentioned, this approach reduces the
size of the uncertainty in the direct amplitude. The effect of absorbing
the order-$\alpha_s$ and the order-$v^2$ contributions in the NRQCD
expansion of $f_V^\parallel$ into an overall factor $f_V$ that is
computed from the quarkonium electronic decay rate corresponds to
correction~(3). Correction~(3) changes the rate for $Z\to J/\psi+\gamma$
by $-13\,\%$ and changes the rate for $Z\to \Upsilon(1S)+\gamma$ by
$-2\,\%$.

As we have mentioned, our result for the relative sign between the direct
and indirect amplitudes disagrees with that in
Ref.~\cite{Huang:2014cxa}. Correction~(4) accounts for the effects of
this change in the relative sign of the indirect amplitude. The
numerical effect of correction~(4) is very small, changing the rates by
only about $2\,\%$, and is insignificant in comparison with the
uncertainties in the rates.

In Ref.~\cite{Huang:2014cxa}, the resummation of logarithms of
$m_Z^2/m_Q^2$ to all orders in $\alpha_s$ was estimated to produce a
$1.5\,\%$ effect in the rate for $Z\to J/\psi +\gamma$. However, we find a
much larger effect, namely, $+18\,\%$. We find that the effect of
the resummation in the rate for $Z\to \Upsilon(1S)+\gamma$ is
$+11\,\%$. Correction~(5) accounts for these resummation 
corrections.

In Ref.~\cite{Huang:2014cxa} the initial scale $\mu_0=m_Q$ was chosen.
As we have explained, we have taken $\mu_0=\overline{m}_Q$ in
order to avoid renormalon ambiguities. We have also replaced $m_Q$
with $\overline{m}_Q$ in the expression for $\phi_V^{\parallel(1)}$ in
Eq.~(\ref{eq:phiV-alphas}). These differences affect the rate for $Z\to
J/\psi +\gamma$ by only $+2\,\%$ and affect the rate for $Z\to
\Upsilon(1S)+\gamma$ by only $+1\,\%$. Correction~(6) accounts for
these differences.

It was claimed in Ref.~\cite{Huang:2014cxa} that only the contributions
of the charm-quark, bottom-quark, and $\tau$-lepton loops are important
in the indirect amplitude. However, we find that these contributions
yield $-43\,\%$ of the real part of the indirect amplitude in the
case of $Z\to J/\psi+\gamma$ and $8\,\%$ of the real part of the indirect
amplitude in the case of $Z\to \Upsilon(1S)+\gamma$.

Our uncertainties are considerably smaller than those in
Ref.~\cite{Huang:2014cxa}. The differences in uncertainties arise from
two principal sources: (1) we have calculated $f_V^\parallel$ from the
leptonic width of the quarkonium, using Eq.~(\ref{EM-width}), instead of
using the perturbative expression in Eq.~(\ref{fV-expn}); and (2) we
have taken into account the known short-distance coefficient $1/5$ for
the order-$v^4$ corrections in estimating the size of these
uncalculated corrections.

\subsection{Comparison with the results from Ref.~[3]}

Our branching fraction for $Z\to J/\psi+\gamma$ differs from that in
Ref.~\cite{Grossmann:2015lea} about $+12\,\%$, which is about
$+2.0\,\sigma$ in the uncertainties of Ref.~\cite{Grossmann:2015lea}.
Our branching fraction for $Z\to \Upsilon(1S)+\gamma$ differs from that
in Ref.~\cite{Grossmann:2015lea} about $-11\,\%$, which is about
$-4.0\,\sigma$ in the uncertainties of Ref.~\cite{Grossmann:2015lea}.

The differences between our results for the central values of the
branching fractions and those of Ref.~\cite{Grossmann:2015lea} arise
primarily because our calculations differ from the calculations in
Ref.~\cite{Grossmann:2015lea} in the following respects: (1) we have
included the nonlogarithmic part of the order-$\alpha_s$ correction to
the LCDA; (2) we have taken  $\mu_0=\overline{m}_Q$ for the initial
scale, instead of $\mu_0=1$~GeV, and we have replaced $m_Q$ with
$\overline{m}_Q$ in the expression for $\phi_V^{\parallel(1)}$ in
Eq.~(\ref{eq:phiV-alphas}); (3) we have used different values of
$\langle v^2\rangle_V$; (4) we have included order-$\alpha_s^2$
contributions to the rate that arise from the absolute square of the
order-$\alpha_s$ correction to the hard-scattering kernel $T_H$; (5) we
have included NLL corrections to the evolution of the LCDA; and (6) we
have included the indirect amplitude.

The effects of these differences on the branching fractions are
tabulated in Table~\ref{tab:gross-uncertainty}. 
As was the case for the corrections to
the calculations in Ref.~\cite{Huang:2014cxa}, the fractional change in
the branching fraction from each correction depends on the order in
which the corrections are incorporated into the calculation. In
Table~\ref{tab:gross-uncertainty}, the fractional changes are computed
by incorporating the corrections in the order (1), (2), (3), (4), (5),
(6). For each quarkonium state, the product of fractional changes gives
the fractional change between our result and that of
Ref.~\cite{Grossmann:2015lea}, aside from some differences of less than
$0.4\,\%$ that arise from small differences in the values that are used
for the Fermi constant, the heavy-quark pole masses, and the decay
constants. As can be seen from Table~\ref{tab:gross-uncertainty}, the
largest correction to the rate for $Z\to J/\psi +\gamma$ arises from
the inclusion of the nonlogarithmic part of the order-$\alpha_s$
correction to the LCDA. This correction is about $+12\,\%$. The largest
correction to the rate for $Z\to \Upsilon(1S) +\gamma$ arises from the
use of a different value of $\langle v^2\rangle_{\Upsilon(1S)}$. This
correction is about $-5\,\%$.
\begin{table}[h]
\begin{center}
\begin{tabular}{lllllll} 
\hline
$\phantom{xx}V\phantom{x}$ & 
\phantom{xx}(1) &
\phantom{xx}(2) &
\phantom{xx}(3) &
\phantom{xx}(4) &
\phantom{xx}(5) &
\phantom{xx}(6) 
\\
\hline\hline
$\phantom{x}J/\psi$\phantom{x} & 
$+11.62\,\%$\phantom{xx}  &
$-0.15\,\%$\phantom{xx} &
$-3.47\,\% $\phantom{xx} &
$+0.68\,\%$\phantom{xx} &
$+2.38\,\%$\phantom{xx} &
$+1.02\,\%$ \phantom{x}
\\
$\phantom{x}\Upsilon(1S)\phantom{xxx}$&
$-3.78\,\%$ &
$-3.50\,\%$ &
$-5.21\,\%$ &
$+0.97\,\%$ &
$+1.81\,\%$ &
$-1.14\,\%$ 
\\
\hline
\end{tabular}
\caption{\label{tab:gross-uncertainty}
The effects on the branching fractions of differences between the
calculations in this work and the calculations in
Ref.~\cite{Grossmann:2015lea}. The corrections (1)--(6) are described in
the text.
}
\end{center}
\end{table}

The uncertainties in the rates that are given in
Ref.~\cite{Grossmann:2015lea} are much smaller than the uncertainties
that we find. In Ref.~\cite{Grossmann:2015lea}, uncertainties from
uncalculated order-$\alpha_s$ corrections are estimated by varying the
hard-scattering scale $\mu$. This approach does not take into account
uncertainties from uncalculated QCD corrections to the LCDA at the
initial scale $\mu_0$ of orders $\alpha_s(\mu_0)$, $\alpha^{2}_s(\mu_0)$,
and $\alpha_s(\mu_0) v^2$. We estimate the relative uncertainties from
the last two of these sources using the formula $\{[C_F
C_A\alpha_s^2(\overline{m}_Q)/\pi^2]^2+[C_F\alpha_s(\overline{m}_Q)
v^2/\pi]^2\}^{1/2}$, which leads to an uncertainty of $8\,\%$ in the case
of $Z\to J/\psi+\gamma$ and an uncertainty of $2.3\,\%$ in the case of
$Z\to \Upsilon(1S)+\gamma$.  Our calculation shows that the
nonlogarithmic correction to the LCDA of order $\alpha_s$, which is not
included in Ref.~\cite{Grossmann:2015lea}, shifts the rate for $Z\to
J/\psi +\gamma$ by about $12\,\%$ and shifts the rate for $Z\to
\Upsilon(1S)+\gamma$ by about $-4\,\%$. In
Ref.~\cite{Grossmann:2015lea}, an uncertainty of about $6\,\%$ is given
for the rate for $Z\to J/\psi +\gamma$ and an uncertainty of about
$3\,\%$ is given for the rate for $Z\to \Upsilon(1S)+\gamma$. Given
the uncertainties from uncalculated corrections of order
$\alpha_s^2(\mu_0)$ and $\alpha_s(\mu_0)v^2$ and the shifts from the
known corrections of order $\alpha_s(\mu_0)$, we believe that the
uncertainties that are given in Ref.~\cite{Grossmann:2015lea} are
underestimates, especially in the case of the rate for $Z\to J/\psi
+\gamma$.

In Ref.~\cite{Grossmann:2015lea}, the order-$v^2$ correction was
computed through the use of a model LCDA whose second $x$ moment is
adjusted to match the second $x$ moment of the actual order-$v^2$
correction. The use of a model LCDA circumvents the difficulties of
divergent eigenvector series that appear in the resummation of
logarithms $m_Z^2/m_Q^2$. However, the choice of the functional form in
the model introduces new uncertainties into the calculation that are not
present in a first-principles calculation, such as the calculation in the 
present paper. In Ref.~\cite{Koenig:2015pha}, a model LCDA with the same
functional form as the model LCDA in Ref.~\cite{Grossmann:2015lea} was
used to compute both the order-$\alpha_s$ and the order-$v^2$ correction
to the LCDA for the process of Higgs-boson decay to a vector quarkonium
plus a photon. It was noted in Ref.~\cite{Bodwin:2016edd}, that, in this
case, the model LCDA does not reproduce the results of the
first-principles calculations of the order-$\alpha_s$ and the
order-$v^2$ corrections accurately. However, we find that, in the case
of the process $Z\to V+\gamma$, the model LCDA {\it does} reproduce the
results of first-principles calculation of the order-$v^2$ correction to
the LCDA reasonably well. The model LCDA result for the order-$v^2$
correction  differs from the first-principles result by $-1.1\,\%$ in
the case of $Z\to J/\psi+\gamma$ and by $+0.8\,\%$ in the case of $Z\to
\Upsilon(nS)+\gamma$. This suggests that the difficulties with the model
LCDA that were noted in Ref.~\cite{Bodwin:2016edd} may arise because of
the incorporation of order-$\alpha_s$ correction to the LCDA into the
model LCDA. We note that the model LCDA contains contributions of order
$v^4$ and higher. As was pointed out in Ref.~\cite{Bodwin:2016edd},
these contributions are incompatible with the relation between the $x$
moments of the LCDA and the NRQCD LDMEs that is given in
Eq.~(\ref{x-moment-LDME}). Apparently, the (incorrect) higher-order
contributions that are contained in the model LCDA are not numerically
significant at the present level of accuracy.

\section{Summary and discussion \label{sec:summary}}

We have presented a calculation of decay rates for the processes $Z\to
V+\gamma$, where $V$ is one of the vector quarkonia $J/\psi$ or
$\Upsilon(nS)$, with $n=1$, $2$, or $3$. Our results for the branching
fractions for  $Z\to V+\gamma$ are given in Table~\ref{tab:num-rates}.
Our calculations contain corrections through relative orders $\alpha_s$
and $v^2$, as well as logarithms of $m_Z^2/m_Q^2$, resummed at NLL
accuracy to all orders in $\alpha_s$. The use of the Abel-Pad\'e method
\cite{Bodwin:2016edd} allows us to compute for the first time the
resummation effects for the order-$\alpha_s$ corrections to the
quarkonium LCDA and to compute from first principles the resummation
effects for the order-$v^2$ corrections to the quarkonium LCDA. The
rates for $Z\to J/\psi+\gamma$ and $Z\to \Upsilon(1S)+\gamma$ have been
computed previously at lower levels of accuracy 
\cite{Huang:2014cxa,Grossmann:2015lea}. Our computations of the
rates for the decays $Z\to \Upsilon(2S)+\gamma$ and $Z\to
\Upsilon(3S)+\gamma$ are new. We have also verified the expressions for
the order-$v^2$ corrections to the decay rate that are given in
Ref.~\cite{Huang:2014cxa}.

Our central values for the branching fractions differ from
those in Ref.~\cite{Huang:2014cxa} by $-10\%$ for the decay 
$Z\to J/\psi+\gamma$ and by $-3\%$ for the decay $Z\to \Upsilon(1S)+\gamma$.
These differences arise
principally for the following reasons: (1) we have corrected the
value for scale of the quarkonium wave function at the origin that was
used in Ref.~\cite{Huang:2014cxa}; (2) we have corrected the value for
the scale of $\alpha_s$ in the order-$\alpha_s$ corrections to the
quarkonium decay constant that was used in Ref.~\cite{Huang:2014cxa};
 (3) in the direct amplitude, we have replaced the nonrelativistic
expansion of $f_V$ [in terms of $\Psi_V(0)$, $\alpha_s$, and $\langle
v^2\rangle$] that was used in Ref.~\cite{Huang:2014cxa} with an
overall factor $f_V$ that is determined from the quarkonium electronic
decay rate; (4) we have included resummations of logarithms of
$m_Z^2/m_Q^2$ in the direct amplitude, whereas such resummations were
not included in the direct amplitude in Ref.~\cite{Huang:2014cxa}. The
individual corrections (1)--(4) are quite large, but they tend to cancel
each other in the rate. We have also found that the sign of the indirect
amplitude, relative to the direct amplitude, is opposite to the sign
that is reported in Ref.~\cite{Huang:2014cxa}. The numerical
consequences of this change in sign are small.

Our central values for the decay rates differ from those in
Ref.~\cite{Grossmann:2015lea} by $+12\,\%$ for the decay $Z\to
J/\psi+\gamma$ and by $-11\,\%$ for the decay $Z\to \Upsilon(1S)+\gamma$.
In the case of the decay $Z\to J/\psi+\gamma$, most of the shift in the
central value occurs because our calculation includes nonlogarithmic
corrections to the LCDA of order $\alpha_s$, while the calculation in
Ref.~\cite{Grossmann:2015lea} does not. In the case of the decay $Z\to
\Upsilon(1S)+\gamma$, the largest difference between our decay rate and
that of  Ref.~\cite{Grossmann:2015lea} occurs because we take the value
of $\langle v^2\rangle_{\Upsilon(1S)}$ from the potential-model
calculation in Ref.~\cite{Chung:2010vz}, while the calculations in
Ref.~\cite{Grossmann:2015lea} make use of an estimate $\langle
v^2\rangle_{\Upsilon(1S)}=0.1$. Other small differences between the
results of our calculations and those of Ref.~\cite{Grossmann:2015lea}
arise for the following reasons: (1) we take the initial scale of the
LCDA to be the heavy-quark $\overline{\rm MS}$ mass, rather than
$1$~GeV; (2) we include the order-$\alpha_s^2$ contribution to the rate
that comes from the absolute square of the order-$\alpha_s$ correction to
the hard-scattering kernel; (3) we resum logarithms of
$m_Z^2/\overline{m}_Q^2$ at NLL accuracy, rather than LL accuracy; and
(4) we include the indirect decay amplitude. We argue that the
choice of the heavy-quark mass as the initial scale of the LCDA is
more appropriate than the choice $1$~GeV because the heavy-quark
mass is the typical scale of perturbative loop corrections to the LCDA.

It is argued in Ref.~\cite{Grossmann:2015lea} that the value of $\langle
v^2\rangle_{\Upsilon(1S)}$ in Ref.~\cite{Chung:2010vz} cannot be correct
because it is negative. However, the minimal-subtraction expression for
$\langle v^2\rangle_{\Upsilon(1S)}$ is obtained by subtracting a power
divergence. Hence, there is no reason that  $\langle
v^2\rangle_{\Upsilon(1S)}$ must be nonnegative. One can see that this is
so by computing, for example, the  minimal-subtraction expression for
$\langle v^2\rangle$ for positronium. In the case of positronium, a full
calculation, including binding effects, can be carried out reliably in
perturbation theory. That computation results in a negative value for
$\langle v^2\rangle$.

The uncertainties in our decay rates are considerably larger than those 
in Ref.~\cite{Grossmann:2015lea}. In  Ref.~\cite{Grossmann:2015lea}, 
uncertainties that arise from uncalculated corrections of higher orders 
in $\alpha_s$ were estimated by varying the hard-scattering scale 
$\mu\sim m_Z$. This procedure does not take into account QCD
corrections to the LCDA, which reside at a scale $\mu_0\sim m_Q$ and 
which were not included in the expression for the amplitude in  
Ref.~\cite{Grossmann:2015lea}. Therefore, we believe that the procedure 
in Ref.~\cite{Grossmann:2015lea} underestimates that uncertainties in 
the rates. 

In Ref.~\cite{Grossmann:2015lea}, the order-$v^2$ correction to the LCDA
were computed by making use of a model for the LCDA whose second $x$
moment, in the narrow-width approximation, agrees with the second $x$
moment of the order-$v^2$ correction to the LCDA. Such a procedure
obviates the use of the Abel-Pad\'e method. However, it
introduces model uncertainties that may not be quantifiable. In
Ref.~\cite{Bodwin:2016edd}, it was found that the use of such a model LCDA
for both the order-$\alpha_s$ and the order-$v^2$ corrections to the
LCDA does not produce accurate results. However, we have found that, when
the model LCDA is used to account only for the order-$v^2$ correction to
the LCDA, it leads to results that differ from our first-principles
calculation only by amounts that are, numerically, of order $v^4$.

The calculations of the decay rate for $Z\to V+\gamma$ in the present paper
improve upon the accuracy of previous theoretical predictions for those
rates and give, we believe, more realistic estimates of the theoretical
uncertainties. Measurements of the decays $Z\to V+\gamma$ are
interesting in their own right as tests of the standard model and as
tests of our understanding of the formation of quarkonium bound states
in hard-scattering processes. However, such measurements are also
important because they can lead to a better understanding of the
experimental difficulties in the observation of quarkonium-plus-photon
final states. That understanding may facilitate the observation of the
rare decays of the Higgs boson to quarkonium-plus-photon final states,
which could yield a first measurement of the Higgs-boson-charm-quark
coupling and alternative measurements of the Higgs-boson-bottom-quark
coupling.

\appendix

\section*{Appendix: Evolution of the LCDA \label{app:evolution}}
The evolution of the LCDA $\phi_V^\parallel(x,\mu)$ is governed by the
Efremov-Radyushkin-Brodsky-Lepage (ERBL) equation
\cite{Efremov:1979qk,Efremov:1978rn,Lepage:1980fj}:
\begin{equation}
\mu^2\frac{\partial}{\partial \mu^2}
 \phi_V^\parallel(x,\mu)
=
\int_0^1 dy\, V_\parallel[x,y;\alpha_s(\mu)]\,
\phi_V^\parallel(y,\mu),
\label{ERBL-evolution}
\end{equation}
where the order-$\alpha_s$ and order-$\alpha_s^2$ contributions to the
ERBL kernel for the longitudinally polarized meson
$V_\parallel[x,y;\alpha_s(\mu)]$ are given in Refs.~\cite{Jia:2008ep}
and \cite{Jones:2007zd}, respectively. The solution of 
Eq.~(\ref{ERBL-evolution}) is given, through NLL order, by 
\cite{Agaev:2010aq}
\begin{equation}
\phi_{n}^\parallel(\mu)|^\textrm{NLL}
=
U_{nk}(\mu,\mu_0)
\phi_n^\parallel(\mu_0),
\end{equation} 
where $U_{nk}(\mu,\mu_0)$ is defined by
\begin{equation}
U_{nk}(\mu,\mu_0)
=
\begin{cases}
E_n^\textrm{NLO}(\mu,\mu_0), & \textrm{if}~k=n,
\\
\frac{\alpha_s(\mu)}{4\pi}
E_n^\textrm{LO}(\mu,\mu_0)d_{nk}(\mu,\mu_0), & \textrm{if}~k<n.
\end{cases}
\end{equation}
Here,
\begin{eqnarray}
E_n^\textrm{LO}(\mu,\mu_0)
&\equiv&
\left[
\frac{\alpha_s(\mu)}{\alpha_s(\mu_0)}
\right]^{\frac{\gamma_n^{\parallel(0)}}{2\beta_0}},
\nonumber \\
E_n^\textrm{NLO}(\mu,\mu_0)
&\equiv&
E_n^\textrm{LO}(\mu,\mu_0)
\left[
1+
\frac{\alpha_s(\mu)-\alpha_s(\mu_0)}{4\pi}
\frac{\gamma_n^{\parallel(1)}\beta_0
-\gamma_n^{\parallel(0)}\beta_1}{2\beta_0^2}
\right].
\end{eqnarray}
The one-loop and two-loop QCD beta-function coefficients are given,
respectively, by
\begin{eqnarray}
\beta_0&\equiv&\frac{11}{3}C_A-\frac{4}{3}T_F n_f,
\nonumber \\
\beta_1&\equiv&\frac{34}{3}C_A^2-\frac{20}{3}C_A T_F n_f
-4C_FT_Fn_f,
\end{eqnarray}
where, as we have already noted, $C_F=(N_c^2-1)/(2N_c)$ and
$C_A=N_c=3$ for color $\textrm{SU}(3)$.  $T_F=1/2$, and $n_f$ is
the number of the active quark flavors. The LO anomalous dimension
$\gamma_n^{\parallel(0)}$ is given by \cite{Ball:1996tb, Jia:2008ep,
Jones:2007zd}
\begin{equation}
\label{eq:gam-n-parallel}
\gamma_n^{\parallel(0)}=
8C_F
\left[
H_{n+1}-\frac{3}{4}-\frac{1}{2(n+1)(n+2)}
\right],
\end{equation}
where
\begin{equation}
H_n=\sum_{j=1}^n\frac{1}{j}
\end{equation}
is the harmonic number.
The NLO anomalous dimension $\gamma_{n-1}^{\parallel(1)}$ is given in 
Ref.~\cite{GonzalezArroyo:1979df} as
\begin{eqnarray}
\gamma_{n-1}^{\parallel(1)}
&=&
\left(C_F^2-\frac{1}{2}C_F C_A\right)
\Bigg\{
16H_n\frac{2n+1}{n^2(n+1)^2}
+16\left[2H_n-\frac{1}{n(n+1)}
\right]
\left(H_n^{(2)}-S_{n/2}^{'(2)}\right)
\nonumber \\
&&\quad\quad\quad\quad\quad\quad\quad\quad
+64\tilde{S}_n+24H_n^{(2)}-3-8S_{n/2}^{'(3)}
-8\frac{3n^3+n^2-1}{n^3(n+1)^3}
-16(-1)^n \frac{2n^2+2n+1}{n^3(n+1)^3}
\Bigg\}
\nonumber \\
&&
+C_FC_A
\Bigg\{
H_n
\left[
\frac{536}{9}+8\frac{2n+1}{n^2(n+1)^2}
\right]
-16H_n H_n^{(2)}
+H_n^{(2)}
\left[
-\frac{52}{3}+\frac{8}{n(n+1)}
\right]
\nonumber \\
&&\quad\quad\quad\quad
-\frac{43}{6}
-4\frac{151n^4+263n^3+97n^2+3n+9}{9n^3(n+1)^3}
\Bigg\}
\nonumber \\
&&
+C_F\frac{n_f}{2}
\Bigg\{
-\frac{160}{9}H_n+\frac{32}{3}H_n^{(2)}
+\frac{4}{3}
+16\frac{11n^2+5n-3}{9n^2(n+1)^2}
\Bigg\},
\end{eqnarray}
where
\begin{eqnarray}
H_{n}^{(k)}&\equiv&\sum_{j=1}^n \frac{1}{j^k},
\quad
\textrm{with}
\quad
H_n^{(1)}\equiv H_{n},
\\
S^{'(k)}_{n/2}&\equiv&
\begin{cases}
\displaystyle
H^{(k)}_{n/2},
& \textrm{if $n$ is even,}
\\
\displaystyle
H^{(k)}_{(n-1)/2},
& \textrm{if $n$ is odd,}
\end{cases}
\\
\tilde{S}_n
&\equiv&
\sum_{j=1}^n \frac{(-1)^j}{j^2}H_j.
\end{eqnarray}
The off-diagonal evolution factor $d_{nk}(\mu,\mu_0)$ is 
\begin{equation}
d_{nk}(\mu,\mu_0)
=
\frac{M_{nk}}
{\gamma_n^{\parallel(0)}-\gamma_k^{\parallel(0)}-2\beta_0}
\left\{
1-\left[
\frac{\alpha_s(\mu)}{\alpha_s(\mu_0)}
\right]^{\frac{\gamma_n^{\parallel(0)}-\gamma_k^{\parallel(0)}
-2\beta_0}{2\beta_0}}
\right\},
\end{equation}
where
\begin{eqnarray}
M_{nk}&=&
\frac{(k+1)(k+2)(k+3)}{(n+1)(n+2)}
(\gamma_n^{\parallel(0)}-\gamma_k^{\parallel(0)})
\left[
\frac{8C_F A_{nk}-\gamma_k^{\parallel(0)}-2\beta_0}
{(n-k)(n+k+3)}
+4C_F\frac{A_{nk}-\psi(n+2)+\psi(1)}{(k+1)(k+2)}
\right],
\nonumber \\
A_{nk}&=&
\psi\left(\frac{n+k+4}{2}\right)-\psi\left(\frac{n-k}{2}\right)
+2\psi(n-k)-\psi(n+2)-\psi(1),
\end{eqnarray}
and $\psi(n)$ is the digamma function. 

\begin{acknowledgments}                                      
We thank Deshan Yang for clarifying several issues with regard to the
formulas in Ref.~\cite{Wang:2013ywc}. We also thank Matthias Neubert
and Matthias K\"onig for a helpful discussion. The work of G.T.B.\ is
supported by the U.S.\ Department of Energy, Division of High Energy
Physics, under Contract No.\ DE-AC02-06CH11357. The work of H.S.C.\ at
CERN is supported by the Korean Research Foundation (KRF) through the
CERN-Korea fellowship program. The work of J.-H.E.\ and J.L.\ was 
supported by the National Research Foundation of Korea (NRF) 
under Contract No. NRF-2017R1E1A1A01074699.
The submitted manuscript has been created in
part by UChicago Argonne, LLC, Operator of Argonne National Laboratory.
Argonne, a U.S.\ Department of Energy Office of Science laboratory, is
operated under Contract No.\ DE-AC02-06CH11357. The U.S.\ Government
retains for itself, and others acting on its behalf, a paid-up
nonexclusive, irrevocable worldwide license in said article to
reproduce, prepare derivative works, distribute copies to the public,
and perform publicly and display publicly, by or on behalf of the
Government.
\end{acknowledgments}


\end{document}